\documentclass[sigconf,nonacm]{acmart}
\AtBeginDocument{%
  }

\setcopyright{acmlicensed}
\copyrightyear{2018}
\acmYear{2018}
\acmDOI{XXXXXXX.XXXXXXX}
\acmConference[Conference acronym 'XX]{Make sure to enter the correct
  conference title from your rights confirmation email}{June 03--05,
  2018}{Woodstock, NY}
\acmISBN{978-1-4503-XXXX-X/2018/06}

\begin{document}

\title{Influence of Agent Number and Behavior on Users' Psychological Responses in VR Evacuation Training}

\author{Naoki Matsumura}
\orcid{0009-0005-4711-7753}
\affiliation{
  \institution{Kansai University}
  \city{Takatsuki}
  \state{Osaka}
  \country{Japan}
}
\email{k452174@kansai-u.ac.jp}

\author{Kensuke Sugihara}
\affiliation{
  \institution{Kansai University}
  \city{Takatsuki}
  \state{Osaka}
  \country{Japan}
}
\email{k631113@kansai-u.ac.jp} 

\author{Tomoko Yonezawa}
\orcid{0000-0001-8226-0815}
\affiliation{
  \institution{Kansai University}
  \city{Takatsuki}
  \state{Osaka}
  \country{Japan}
}
\email{yone@kansai-u.ac.jp}

\renewcommand{\shortauthors}{Naoki Matsumura, Kensuke Sugihara, and Tomoko Yonezawa}

\begin{abstract}
This study aims to build effective drill software by combining a virtual simulation of a disaster with multiple agents capable of expressing emotions. We evaluated how the number of agents and their emotional and behavioral actions during evacuation influence users in a virtual space. The results showed that users tended to act more confidently when many agents evacuated with them, and that the evacuation route chosen by the agents influenced the users' decision-making. These findings indicate that users' evacuation behavior is shaped not only by the environment itself but also by the collective behavior of surrounding agents, underscoring the need to consider how agent behavior affects users' psychological states during training.
\end{abstract}


  


\settopmatter{printacmref=false}

\maketitle

\section{Introduction}
In recent years, various types of disasters—including earthquakes, fires, and floods—have occurred frequently around the world. Consequently, the importance of disaster preparedness training aimed at acquiring rapid and appropriate evacuation behavior has increased. Conventional evacuation drills are typically designed to confirm evacuation routes and procedural steps. However, prior studies have shown that actual evacuation behavior is strongly influenced by psychological factors such as intense fear, stress, and the actions of surrounding people \cite{helbing2013}. In particular, conformity bias, in which individuals follow the behavior of others, has been identified as a major cause of crowd accidents and failures in fire evacuation \cite{cialdini2004}.

To address these psychological factors, VR (Virtual Reality)-based disaster experience systems have attracted increasing attention \cite{nakahara2026}. Moreover, recent studies have simulated crowd movement and social influence by placing multiple agents within virtual environments, enabling researchers to observe how evacuees are affected by the behavior of others \cite{kasahara2026, zheng2026}. While many of these studies primarily focus on reproducing crowd behavior or enhancing the realism of evacuation simulations, relatively few attempts have been made to use agent-related parameters—such as the number of agents or their behavioral tendencies—to evaluate or improve psychological aspects of evacuation training.

In this study, we manipulate the number of agents evacuating simultaneously with the user and the ratio of their route choices within a VR environment. We then examine how these factors influence users' psychological indicators—including fear, sense of safety, confidence, and hesitation—as well as their evacuation success rate. Through this investigation, we aim to provide insights for designing VR-based disaster education systems that incorporate multiple agents to promote awareness of psychological influences such as conformity bias and to support calm, autonomous decision-making during emergencies.

\section{Related Research}
\subsection{VR-based Disaster Preparedness Training}
In recent years, many systems have been proposed that use VR to reproduce disaster situations such as fires and earthquakes, with the aim of improving the understanding of evacuation behavior and enhancing disaster-prevention education \cite{hung2025}. VR has been recognized as a promising tool for strengthening the effectiveness of disaster education \cite{kaggwa2025}. These studies leverage the immersive nature of VR to present visual and auditory information that is difficult to convey through conventional teaching materials, thereby supporting experiential learning of evacuation behavior.

Research that reproduces route-selection and wayfinding behavior in architectural spaces has evaluated how environmental factors such as visibility, placement of emergency lights, and the behavior of others influence evacuees' route recognition and decision-making. These findings demonstrate that VR is an effective method for observing cognitive processes during evacuation \cite{arias2018}.

Studies examining the educational effectiveness of VR-based fire evacuation training have reported that high-fidelity disaster experiences, including spreading flames, smoke, and alarm sounds, significantly improve participants' understanding of evacuation procedures and recognition of hazardous situations compared with traditional video materials \cite{muffato2026}. Furthermore, research that realistically reproduces disaster conditions involving limited visibility due to flames and smoke has shown that the immersive nature of VR enhances the realism of decision-making during evacuation \cite{han2026}.

Many existing studies have aimed to realistically reproduce disaster scenes in VR, using immersive experiences to enhance understanding of evacuation behavior and improve the effectiveness of disaster-prevention education. However, it has been reported that panic and collective psychological effects can impair calm decision-making during actual emergencies \cite{lamrhary2026}. Therefore, simply increasing immersion may not be sufficient for training individuals to make appropriate decisions in real disaster situations.

\subsection{VR-based Psychological Training}
Evacuation behavior during disasters is strongly influenced by psychological factors such as normalcy bias and conformity bias, which can impede rational decision-making \cite{yamori2008,kinateder2015}. These cognitive biases may lead to delayed evacuation initiation or inappropriate route choices, highlighting the need for training that helps individuals understand and resist such biases.

Research using multi-user immersive VR environments to analyze crowd evacuation under high-stress conditions has shown that evacuees are strongly influenced by the actions of surrounding individuals, resulting in herding behavior and crowding toward specific routes \cite{moussaid2016}. Similarly, studies analyzing how virtual guides placed within VR environments influence evacuees' decision-making have shown that evacuees select routes by considering the alignment between guide instructions and crowd movements, as well as spatial and social cues \cite{cao2025}.
VR experiments focusing on tsunami evacuation have further revealed that the presence and behavior of others, as well as environmental conditions, influence protective actions and risk perception \cite{navarro2026}. These findings indicate that social influences observed in real-world crowd evacuations can also be reproduced within VR environments.

Several studies have incorporated crowd psychology into VR-based disaster training. For example, VR evacuation training systems have been developed that intentionally induce conformity bias and panic by manipulating NPC behavior, allowing participants to experience inappropriate actions such as following crowds running toward incorrect directions \cite{kasahara2026}. Additionally, VR serious games designed to induce cognitive biases during disasters have demonstrated the effectiveness of metacognitive training that helps participants understand how normalcy bias and conformity bias influence decision-making \cite{arakawa2026}.

These studies collectively demonstrate the potential of VR for experiential training that addresses psychological factors affecting evacuation behavior. However, it remains unclear how the number of agents and the distribution of their route choices within VR environments should be manipulated to effectively train users' psychological responses and decision-making.
Therefore, this study focuses on the number of agents and the ratio of their evacuation route choices in VR environments, and evaluates how these factors influence users' psychological states and evacuation behavior.

\section{Proposal System}
\subsection{System Overview}
This study proposes an immersive disaster preparedness training system using an HMD. The system aims to train users to maintain composure and take proactive actions during a disaster--without succumbing to conformity bias--by conducting evacuation drills alongside multiple agents that exhibit panic behaviors, such as screaming and acting in a flustered manner. The proposed system consists of a PC running Windows 10 and a \href{https://www.vive.com/jp/product/vive-pro-eye/overview/}{VIVE Pro Eye} headset. Unity was used to implement the virtual environment, control agent behavior, and create the evacuation drill content.

\begin{figure}[t]
\centering
\scalebox{0.7}{\includegraphics[clip, width=\hsize]{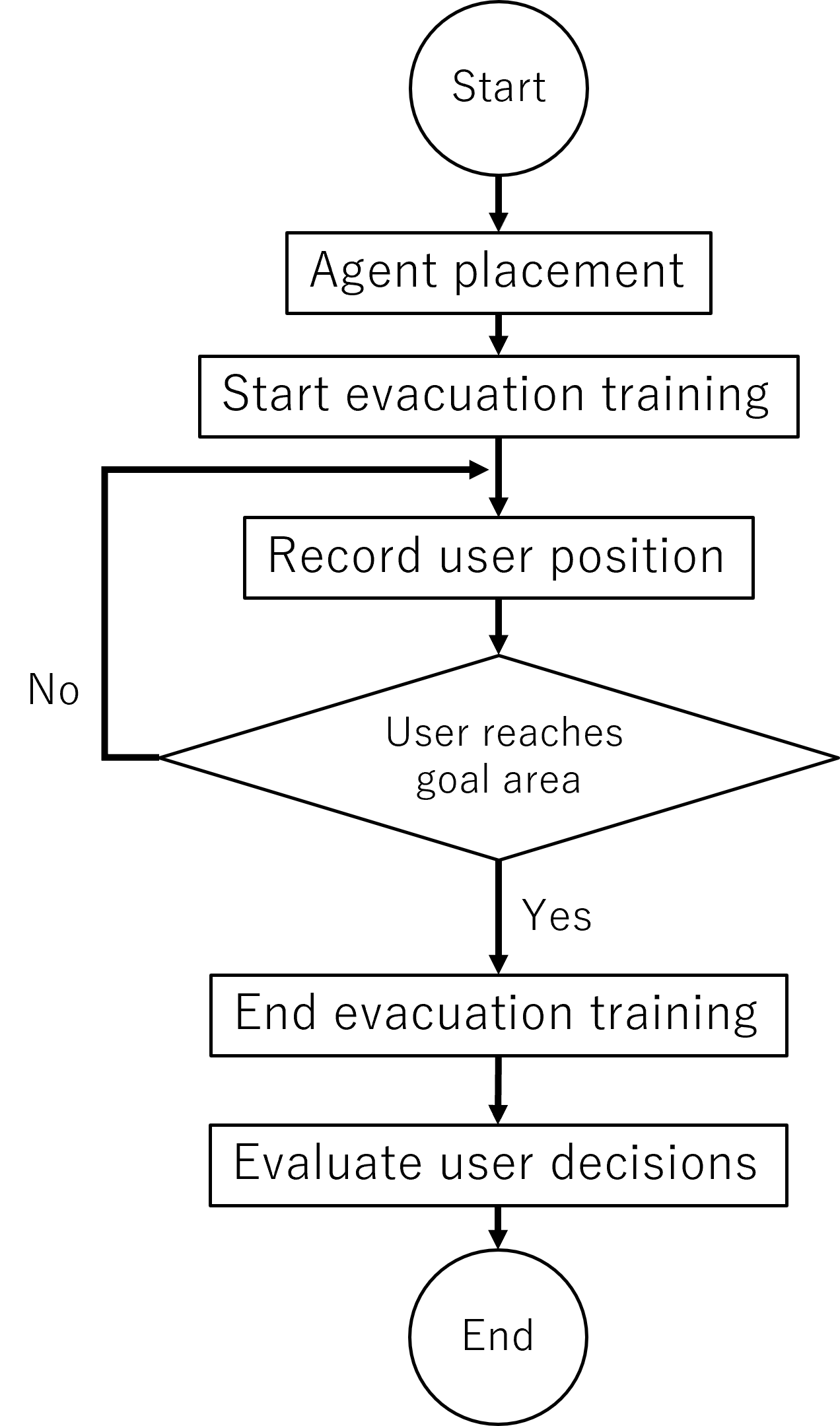}}
\caption{System flow}
\label{system_flow}
\end{figure}

Figure \ref{system_flow} illustrates the system workflow. Upon system startup, agents are positioned in the environment. When the drill begins, the agents start moving, following pre-set routes to evacuate. Users navigate the virtual space by turning their heads in the direction of travel and operating controller buttons. The drill concludes when the user reaches an exit. Throughout the drill, the user's coordinates within the virtual space are recorded every 500 milliseconds; upon completion, this data is used to evaluate whether the user selected an appropriate evacuation route.

\subsection{Agent Behavior}
\begin{figure}[t]
\centering
\scalebox{0.8}{\includegraphics[clip, width=\hsize]{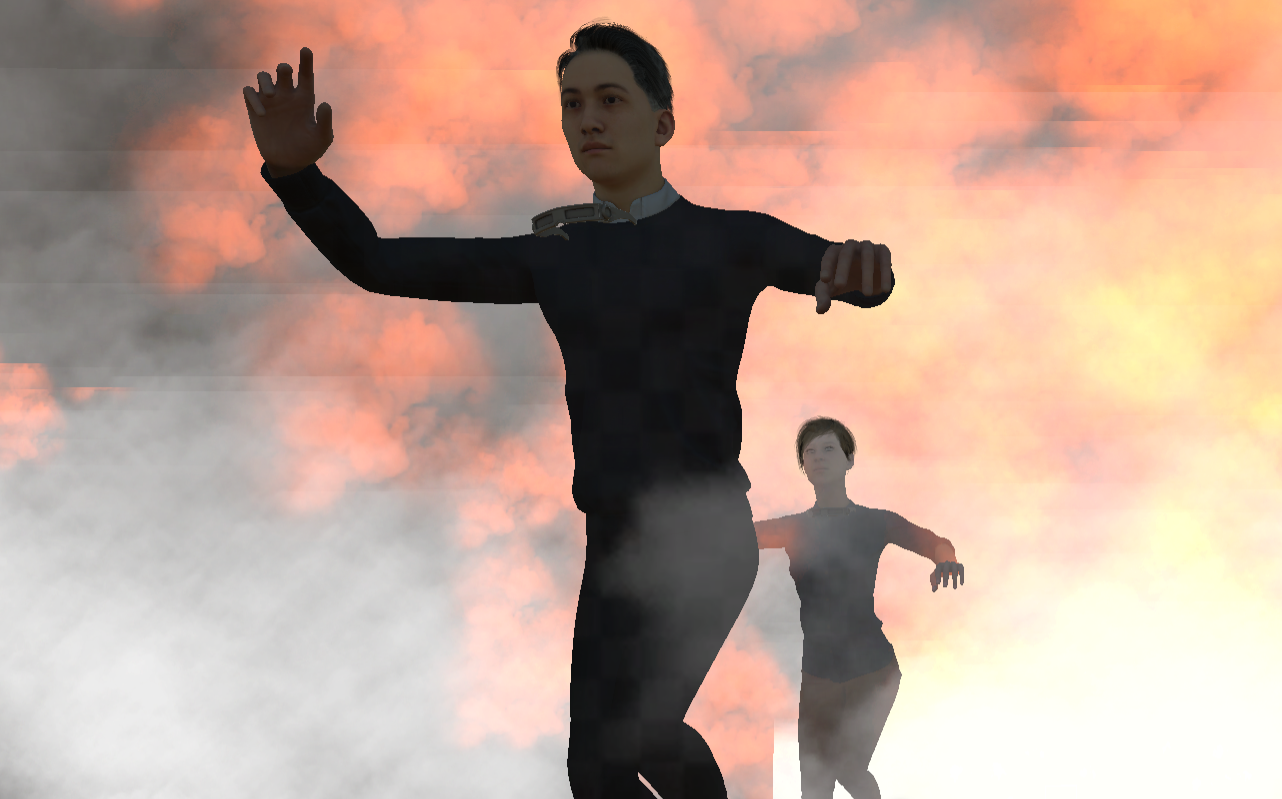}}
\caption{Panicked agent}
\label{panic_agent}
\end{figure}

In the proposed system, multiple agents exhibiting fearful emotions--such as panic--are deployed, and their selection of evacuation routes is controlled. Standard running animations are applied to agents in a calm emotional state, while animations depicting agents running with both hands raised and screaming are applied to those in a state of panic (Figure \ref{panic_agent}). The number and behavior of these agents can be determined at runtime. Each agent is implemented using Unity's NavMesh Agent component to automatically select a path toward pre-configured decision points on the map.

\begin{figure}[t]
\centering
\scalebox{0.8}{\includegraphics[clip, width=\hsize]{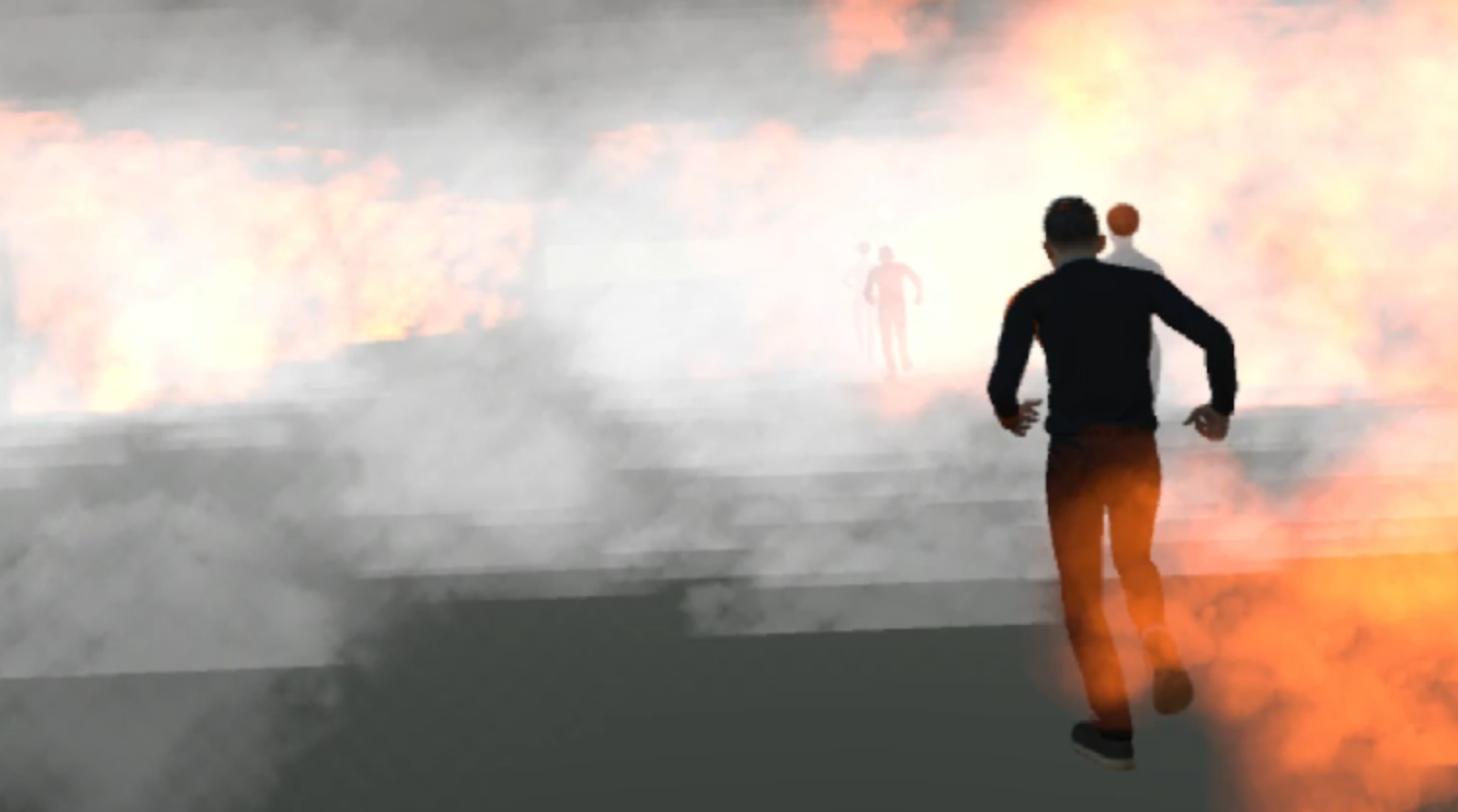}}
\caption{Evacuation drill scene}
\label{evacuation_drill_scene}
\end{figure}

\section{Experiments}
We examined how the behaviors of virtual agents, implemented to heighten participants' sense of presence and fear during disaster evacuation training in a virtual environment, as well as the proportion of agents exhibiting each behavior, influence participants' fear levels and evacuation success rates. The experiment involved 20 participants (18 male and 2 female), aged 21 to 23, all of whom provided informed consent regarding the experimental procedures and the handling of their personal information and rights.

\subsection{System Configuration for the Experiment}
\begin{figure}[t]
\centering
\scalebox{0.8}{\includegraphics[clip, width=\hsize]{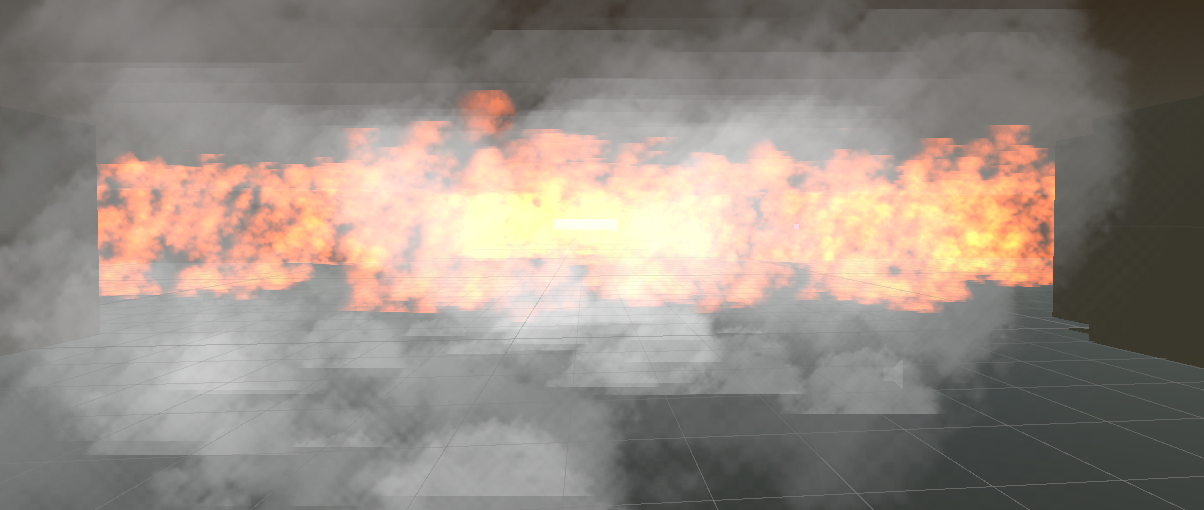}}
\caption{Appearance of an impassable route}
\label{fire_route}
\end{figure}

In the experiment, we simplified the evacuation route options; at a crossroads, the scenario presented a situation where proceeding straight ahead was clearly blocked by visible flames, requiring a choice between turning right or left (Figure \ref{fire_route}). This setup ensures that the appropriateness of the route chosen by the agents--whether or not it constitutes a valid evacuation path--is easily understood.

\subsection{Experiment 1: Effect of the number of agents on evacuation behavior.}

\noindent\textbf{Purpose: } Experiment 1 aims to investigate how the presence and number of agents accompanying the user during evacuation behavior affect the user's sense of security and fear. In particular, this experiment serves as a preliminary study to determine whether the presence or number of agents has any impact.\\

\noindent\textbf{Experimental Hypotheses: }
\begin{itemize}
\item \textbf{Hypothesis 0}: The presence of an agent increases the sense of security.
\item \textbf{Hypothesis 1}: The absence of an agent increases the sense of fear.
\item \textbf{Hypothesis 2}: A larger number of agents increases the sense of security.
\item \textbf{Hypothesis 3}: A smaller number of agents increases the sense of fear.
\end{itemize}

\noindent\textbf{Conditions: }
To simulate the group psychological states associated with varying numbers of agents (including the absence thereof), we established a within-subjects experimental design with one factor and five conditions. We set the maximum number of agents at 20--considering the performance limits of the system running on Unity--and defined the number of agents as Factor A (A1: 0 agents, A2: 1 agent, A3: 3 agents, A4: 8 agents, A5: 20 agents). Furthermore, to isolate the emotional impact of the number of agents, participants were required to follow the exact same route as the agents in conditions where agents were present, whereas they could choose any route in conditions where no agents were present. The presentation order of the experimental conditions was determined using a Latin square design \cite{Grant1948} to ensure counterbalancing.\\

\noindent\textbf{Procedure: }
Participants were instructed on how to wear the HMD and operate the system; to minimize the effects of learning or familiarity with the controls, they were given the opportunity to practice in a virtual environment. Subsequently, they experienced the experimental system under each condition and answered a set of questions after each trial.\\

\noindent\textbf{Evaluation Items: }
We established the following questionnaire items to assess the sense of security and fear experienced during the virtual evacuation drill. Participants rated these items on a scale of 0 to 99 (0: not at all, 99: very strongly) using the Visual Analog Scale (VAS) method.

\begin{enumerate}
\item[Q1.] I felt a sense of urgency during the evacuation drill.
\item[Q2.] I was able to act calmly during the evacuation drill.
\item[Q3.] I felt fear while evacuating.
\item[Q4.] I was able to evacuate with a sense of security.
\item[Q5.] I felt confident in my actions.
\item[Q6.] I felt that my evacuation route was correct.
\end{enumerate}

\begin{figure}[t]
\centering
\scalebox{0.9}{\includegraphics[clip, width=\hsize]{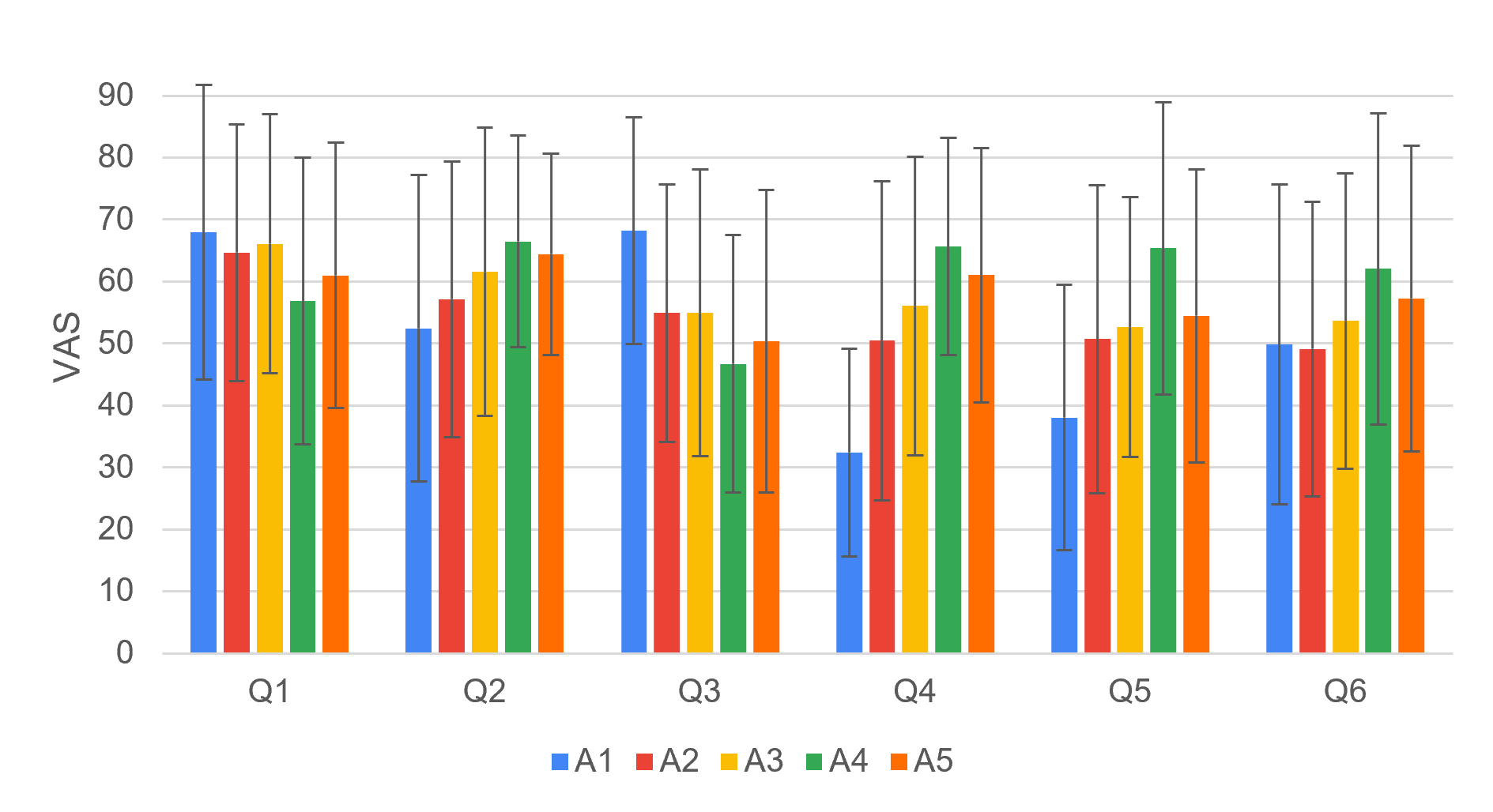}}
\caption{VAS scores in Experiment 1}
\label{m_sd_exp1}
\end{figure}

\noindent\textbf{Results: }
Analysis of variance with a significance level of $\alpha = 0.05$ was conducted for each questionnaire item. The Ryan method was used for multiple comparisons. Figure \ref{m_sd_exp1} presents the mean values and standard deviations for all items, and Table \ref{ana_e1} summarizes the ANOVA results.

Analysis of Q3 revealed significant differences among conditions A1, A4, and A5. Multiple comparisons showed significant differences between A1 and A4, and between A1 and A5, with A1 receiving higher ratings than A4 and A5. In other words, participants felt more fear when no agents evacuated along the same route. This result supports Hypothesis 1. Conversely, no significant differences were found among the conditions where one or more agents were present. Significant differences were also observed for Q4 . Multiple comparisons revealed significant differences between A1 and each of the other conditions (A2, A3, A4, and A5), with the mean ratings for A2, A3, A4, and A5 being higher than that for A1. Participants felt more secure when evacuating alongside one or more agents. Significant differences were also found for Q5 . Multiple comparisons showed a significant difference between A1 and A4, with the mean rating for A4 being higher than that for A1. Participants acted with greater confidence when eight agents accompanied them compared to when they were alone. These results support Hypothesis 0.

Thus, while significant differences were observed between A1 and the other conditions, no significant differences were found among the conditions other than A1; consequently, no effect resulting from an increase in the number of agents was confirmed. Therefore, it is concluded that Hypotheses 2 and 3 were not supported under the experimental settings of this system.

\begin{table}[t]
\caption{ANOVA results for Experiment 1}
\label{ana_e1}
\centering
\scalebox{0.75}{
\begin{tabular}{l|rrl} \toprule
& \multicolumn{2}{l}{A} & \\
Question & \multicolumn{1}{l}{$F$} & \multicolumn{1}{l}{$p$} & Comparison \\ \midrule
Q1 & 1.227 & .307\:\: & --\\
Q2 & 1.957 & .110\:\: & --\\
Q3 & 3.625 & .009$^\ast$ & \{A4, A5\} $\textless$ A1 \\
Q4 & 9.199 & $\textless .001^\ast$ & A1 $\textless$ \{A2, A3, A4, A5\} \\
Q5 & 3.873 & .006$^\ast$ & A1 $\textless$ A4 \\
Q6 & 1.021 & .402\:\: & --\\ \bottomrule
\multicolumn{4}{l}{$Note. \ast ... p \textless .05$}
\end{tabular}
}
\end{table}

\subsection{Experiment 2: Effect of the Ratio of Participating Individuals}

\noindent\textbf{Purpose: }
This experiment investigates how differences in the evacuation route choices made by agents affect participants' feelings of fear, immersion, and reassurance, as well as their sense of confusion and evacuation success rate.\\

\noindent\textbf{Experimental Hypotheses: }
\begin{itemize}
\item \textbf{Hypothesis 4}: The presence of an agent increases the sense of security.
\item \textbf{Hypothesis 5}: The absence of an agent increases the sense of fear.
\end{itemize}

\noindent\textbf{Conditions: }
In Experiment 2, a within-subjects experimental design with one factor and five conditions was established. The number of agents was set to eight, and the proportion of agents choosing the appropriate route served as Factor B (agent behavior ratio: B1: 0\%, B2: 25\%, B3: 50\%, B4: 75\%, B5: 100\%). It was hypothesized that the route chosen by the user would be influenced by the proportion of agents selecting the appropriate evacuation route. Evacuation success was determined based on whether the user chose a route with many or few fire locations, which had been pre-positioned at various points along the evacuation paths. The presentation order of the experimental conditions was determined using the Latin square method \cite{Grant1948} to ensure counterbalancing.\\

\noindent\textbf{Procedure: }
Experiment 2 followed the same procedure as Experiment 1. Additionally, in Experiment 2, the evacuation success rate is recorded for each condition as the percentage of participants who chose the correct route.\\

\noindent\textbf{Evaluation Items: }
For this experiment, questions regarding feelings of fear, reassurance, and confusion were established. Participants rated these items on a scale of 0 to 99 (0: do not feel this way at all; 99: feel this way very strongly) using the VAS method.

\begin{enumerate}
\item[Q1.] I felt a sense of urgency during the evacuation drill.
\item[Q2.] I felt fear while evacuating.
\item[Q3.] I felt a sense of panic or haste while evacuating.
\item[Q4.] I felt confident in my actions.
\item[Q5.] I was concerned about the agents' actions.
\item[Q6.] I was unsure about which evacuation route to choose.
\item[Q7.] I felt that the evacuation route I chose was correct.
\item[Q8.] I was concerned about the ratio of evacuees.
\end{enumerate}

\begin{figure}[t]
\centering
\scalebox{0.9}{\includegraphics[clip, width=\hsize]{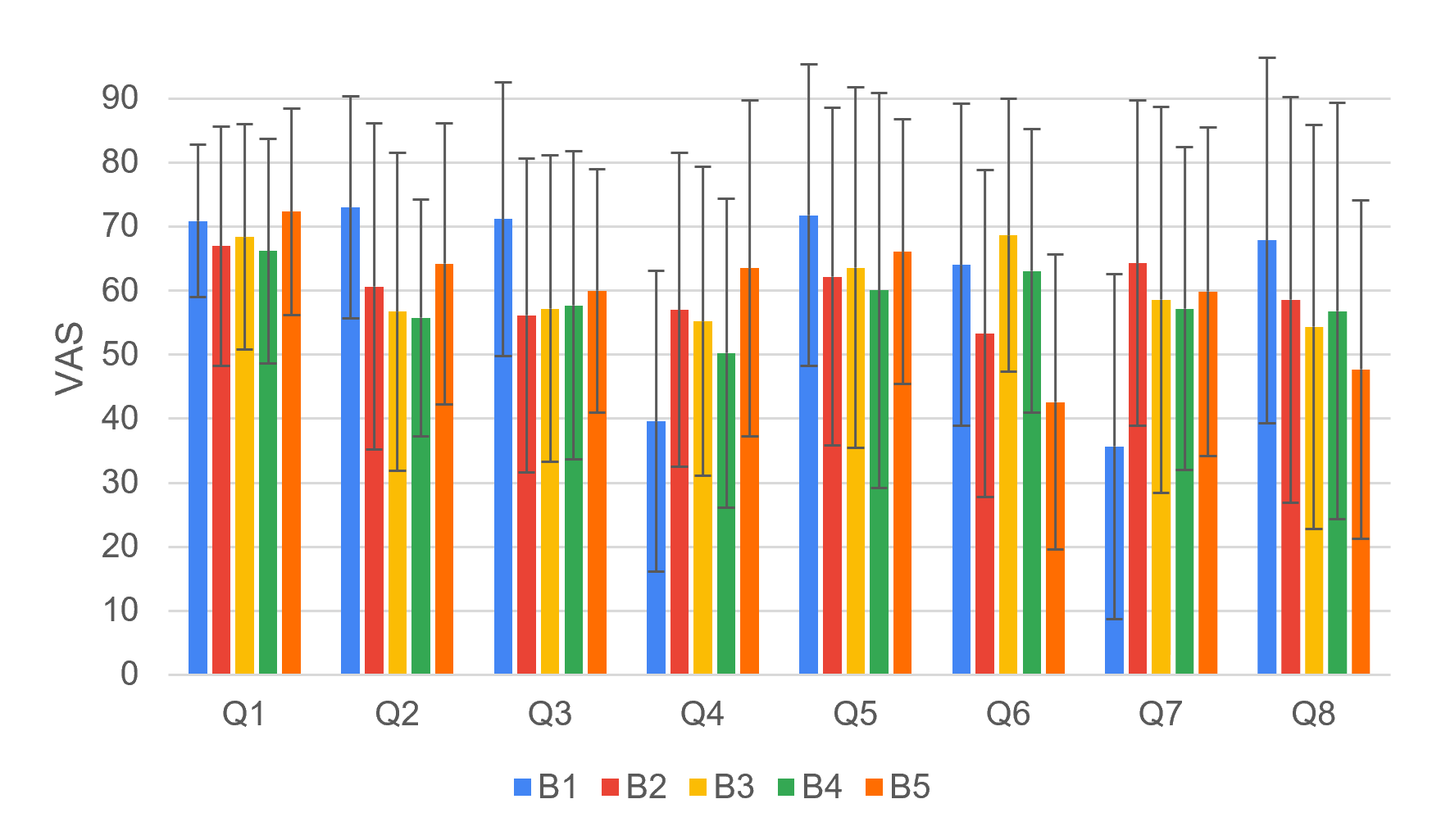}}
\caption{VAS scores in Experiment 2}
\label{m_sd_exp2}
\end{figure}

\noindent\textbf{Results: }
Analysis of variance with a significance level of $\alpha = 0.05$ was conducted for each questionnaire item. The Ryan method was used for multiple comparisons. Figure \ref{m_sd_exp2} presents the mean values and standard deviations for all items, and Table \ref{ana_e2} summarizes the ANOVA results.

First, regarding confidence in their own evacuation behavior, significant differences were found for evaluation items Q4 and Q7. Multiple comparisons revealed a significant difference for Q4 between conditions B1 and B5, with the mean score for B5 being higher than that for B1. For Q7, significant differences were found between condition B1 and conditions B2, B3, B4, and B5; the mean scores for the latter conditions were higher than that for B1. In short, condition B1 yielded the lowest values for both Q4 and Q7. When all agents chose an incorrect route, participants lacked confidence in their own evacuation route choices and rated the correctness of their decisions lower. Thus, it appears that the agents' behavior induced hesitation and anxiety in the participants' decision-making, partially supporting Hypothesis 4. However, significant differences were not confirmed between all levels regarding the continuous changes in selection rates.

A significant difference was also found between conditions for Q6. Multiple comparisons showed significant differences between condition B5 and conditions B1, B3, and B4; the mean scores for B1, B3, and B4 were higher than that for B5, while the value for B2 fell between the mean for B1--B3 and the value for B5. In other words, user hesitation was low when the probability of agents selecting the appropriate evacuation route was 100\%, whereas hesitation increased when choices were dispersed or when all agents selected inappropriate routes.

Regarding evaluations related to a sense of urgency or fear, no significant differences were found for Q1 and Q3. Although a significant difference was observed for Q2, no significant differences were found in the multiple comparisons. Based on these results, Hypothesis 5 was not supported.

Table \ref{esc_suc_2} shows the evacuation success rates of participants in Experiment 2. The success rate was highest for condition B5 and lowest for condition B1. Thus, the participants' evacuation behavior was influenced by the evacuation behavior of the agents. 

\begin{table}[t]
\caption{ANOVA results for Experiment 2}
\label{ana_e2}
\centering
\scalebox{0.75}{
\begin{tabular}{l|rrl} \toprule
& \multicolumn{2}{l}{B} & \\
Question & \multicolumn{1}{l}{$F$} & \multicolumn{1}{l}{$p$} & Comparison\\ \midrule
Q1 & 0.976 & .426\:\: & --\\
Q2 & 2.676 & .038$^\ast$ & --\\
Q3 & 1.555 & .195\:\: & --\\
Q4 & 3.173 & .018$^\ast$ & B1 $\textless$ B5 \\
Q5 & 0.866 & .488\:\: & --\\
Q6 & 3.975 & .006$^\ast$ & B5 $\textless$ \{B1, B3, B4\} \\
Q7 & 3.430 & .012$^\ast$ & B1 $\textless$ \{B2, B3, B4, B5\} \\
Q8 & 1.591 & .185\:\: & --\\ \bottomrule
\multicolumn{4}{l}{$Note. \ast ... p \textless .05$}
\end{tabular}
}
\end{table}

\begin{table}[t]
\caption{Evacuation success rate in Experiment 2}
\label{esc_suc_2}
\centering
\begin{tabular}{l|lllll}
\toprule
Condition & B1 & B2 & B3 & B4 & B5 \\ \midrule
Success rate & 0.10 & 0.55 & 0.55 & 0.65 & 0.80 \\ \bottomrule
\end{tabular}
\end{table}

\subsection{Experiment 3: Effects of Fear on Groups and Choices}

\noindent\textbf{Purpose: }
In this experiment, we investigate whether the fear expressed by virtual agents and their route-selection behaviors in a VR disaster scenario actually influence participants' fear levels and related responses.\\

\noindent\textbf{Experimental Hypotheses: }
\begin{itemize}
\item \textbf{Hypothesis 6}: Combinations of behaviors and emotions among multiple agents influence the user's impressions.
\end{itemize}

\begin{figure}[t]
\centering
\scalebox{0.9}{\includegraphics[clip, width=\hsize]{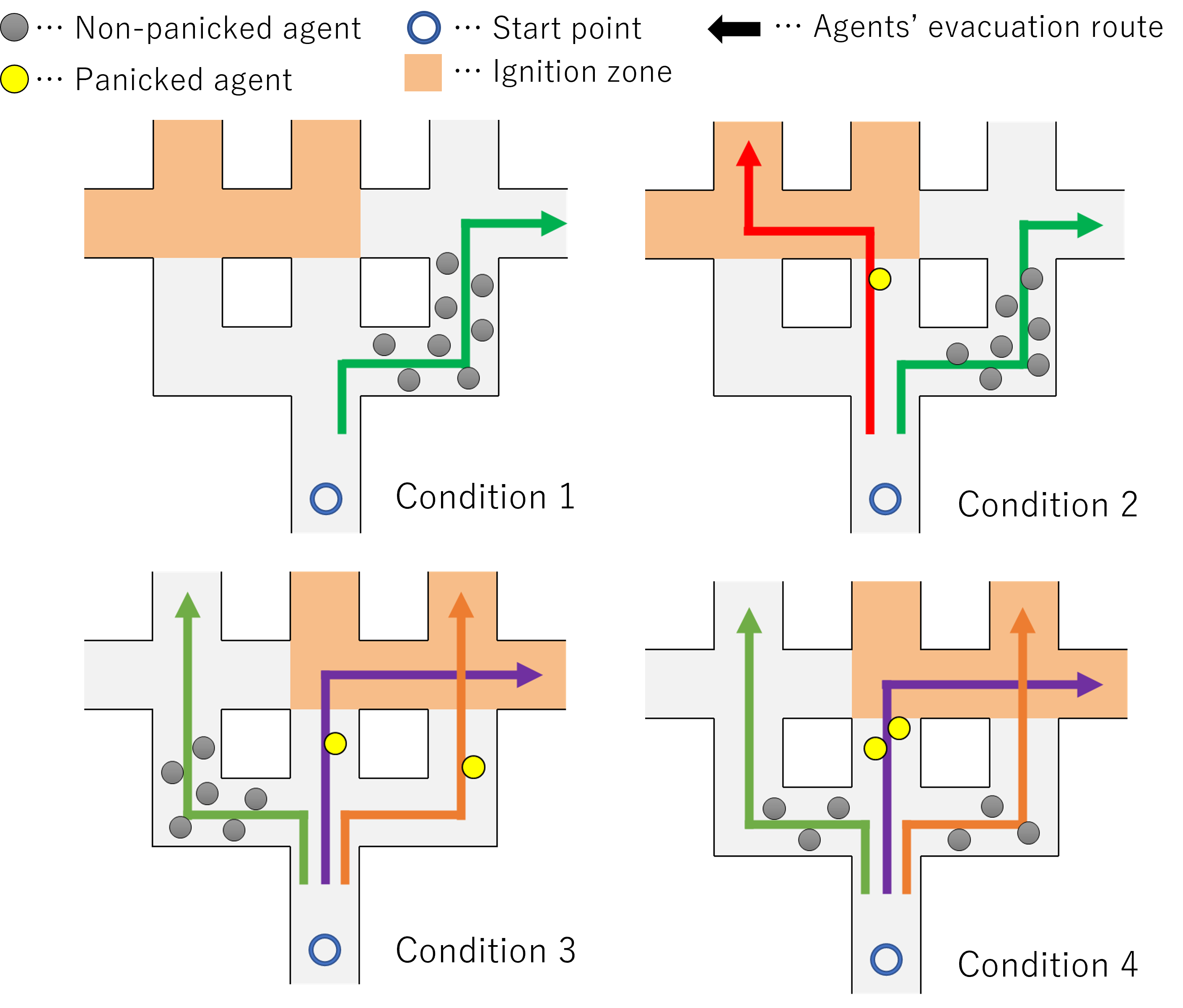}}
\caption{Experimental conditions in Experiment 3}
\label{exp3_condition}
\end{figure}

\noindent\textbf{Conditions: }
In this experiment, the number of agents was fixed at eight, and four experimental conditions were created by varying the presence of emotional expressions and the diversity of route-selection behaviors, as shown in Figure \ref{exp3_condition}. The study employed a within-subjects design with one factor and four conditions. As in Experiment 2, the success or failure of participants' evacuation attempts during the simulated experience was evaluated using the same criteria. Agents with strong emotional expressions were configured to choose inappropriate routes. The order in which the experimental conditions were presented was determined using a Latin square design \cite{Grant1948} to ensure counterbalancing.\\

\noindent\textbf{Procedure: }
Experiment 3 was conducted using the same procedure as Experiment 2. Additionally, the evacuation success rate of the participants in each trial was recorded.\\

\noindent\textbf{Evaluation Items: }
The evaluation items for Experiment 3 are as follows. As in Experiment 2, subjective ratings were assessed using the VAS method, treating the values as interval scales.
\begin{enumerate}
\item[Q1.] Felt like a realistic evacuation drill
\item[Q2.] Felt fear while evacuating
\item[Q3.] Felt confident in my own actions
\item[Q4.] Felt flustered/panicked while evacuating
\item[Q5.] Paid attention to the agents' actions
\item[Q6.] Hesitated in deciding the evacuation route
\item[Q7.] The agents' actions evoked a sense of fear
\end{enumerate}

\begin{figure}[t]
\centering
\scalebox{0.9}{\includegraphics[clip, width=\hsize]{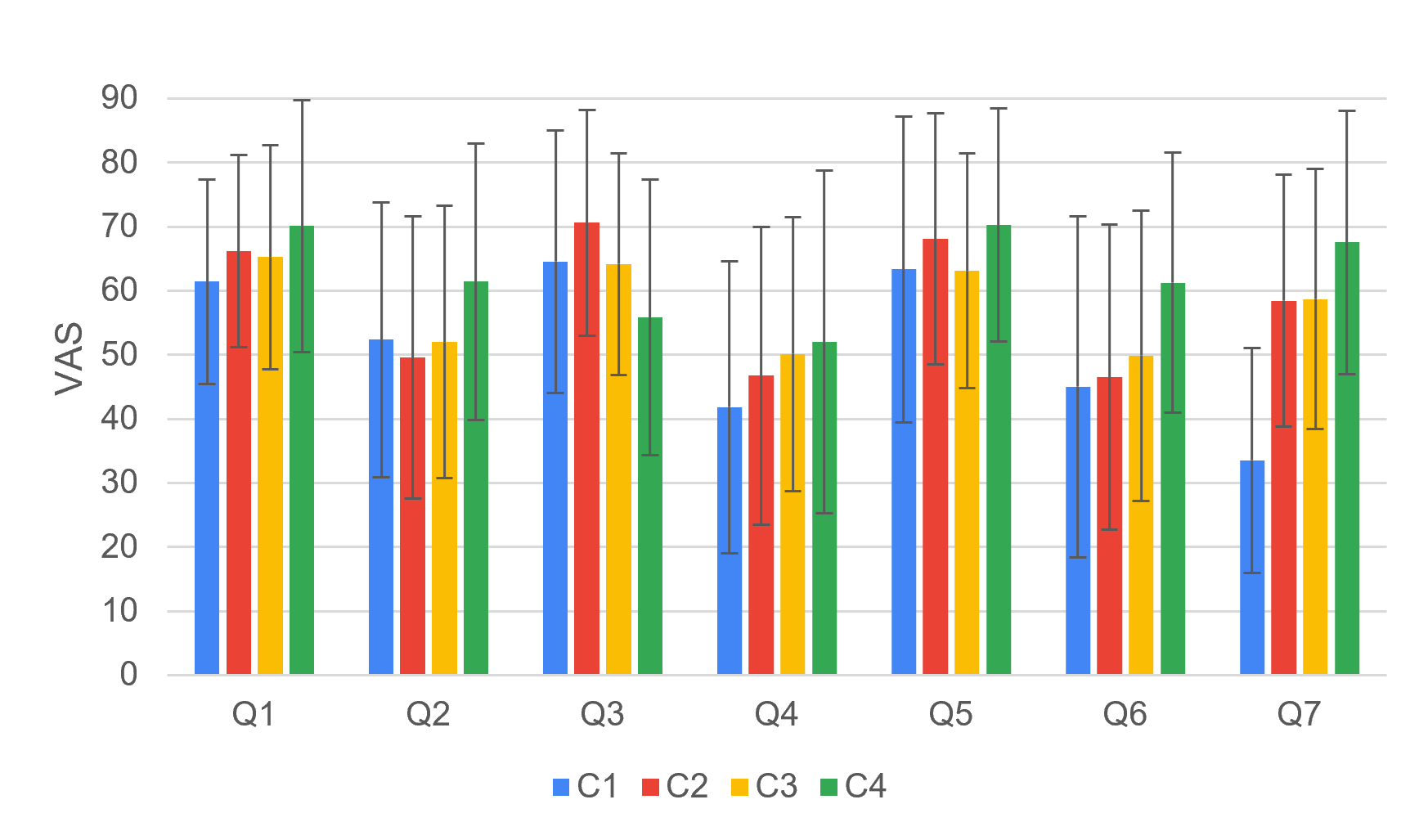}}
\caption{VAS scores in Experiment 3}
\label{m_sd_exp3}
\end{figure}

\noindent\textbf{Results: }
Analysis of variance with a significance level of $\alpha = 0.05$ was conducted for each questionnaire item. The Ryan method was used for multiple comparisons. Figure \ref{m_sd_exp3} presents the mean values and standard deviations for all items, and Table \ref{ana_e3} summarizes the ANOVA results.

A significant difference was found for Q7. The results of the multiple comparison indicated significant differences between C1 and C2, C3, and C4. These findings suggest that a group of agents without variability in behavior or emotional expression may reduce participants' perception of fear. Although Q3 showed only a marginal trend toward significance, the condition with both behavioral and emotional variability (C4) again produced the lowest scores. No significant differences were observed for the remaining items. Regarding Hypothesis 6, these results indicate that variability in agents' behaviors and emotional expressions may influence participants' responses during evacuation. The evacuation success rate (Table \ref{esc_suc_3}) also showed a marked decrease in C4, where variability was highest.

\begin{table}[t]
\caption{ANOVA results for Experiment 3}
\label{ana_e3}
\centering
\scalebox{0.75}{
\begin{tabular}{l|rrl} \toprule
& \multicolumn{2}{l}{C} & \\
Question & \multicolumn{1}{l}{$F$} & \multicolumn{1}{l}{$p$} & Comparison \\ \midrule
Q1 & 1.873 & .144\:\: & --\\
Q2 & 1.711 & .175\:\: & --\\
Q3 & 2.536 & .068$^+$ & --\\
Q4 & 0.951 & .422\:\: & --\\
Q5 & 0.965 & .416\:\: & --\\
Q6 & 1.967 & .129\:\: & --\\
Q7 & 14.303 & $\textless .001^\ast$ & C1 $\textless$ \{C2, C3, C4\} \\ \bottomrule
\multicolumn{4}{l}{$Note.  + ... p \textless .10,   \ast ... p \textless .05$}
\end{tabular}
}
\end{table}

\begin{table}[t]
\caption{Evacuation success rate in Experiment 3}
\label{esc_suc_3}
\centering
\begin{tabular}{l|llll}
\toprule
Condition & C1 & C2 & C3 & C4 \\ \midrule
Success rate & 0.90 & 1.00 & 0.90 & 0.50 \\ \bottomrule
\end{tabular}
\end{table}

\section{Discussion}
\subsection{Results and Possibility}
In Experiment 1, we found that participants experienced stronger fear during the evacuation training when no agents accompanied them, compared with conditions in which 8 or 20 agents were present. This suggests that adjusting the number of agents may allow evacuation training systems to intentionally evoke users’ anxiety or fear during evacuation. However, the possibility that fear was influenced simply by evacuating in an unfamiliar environment was not examined. In the no-agent condition, participants had to determine the evacuation route themselves in an unfamiliar environment, whereas in the agent-present conditions, they could reach the exit by following the agents. Therefore, the observed increase in fear may have been influenced not only by the number of agents but also by the cognitive burden and uncertainty involved in route selection when the exit location was unclear. The extent to which agents' behaviors and participants' familiarity with the evacuation environment affect fear and evacuation behavior remains an important topic for future investigation.

In Experiment 2, results showed that when all agents selected the appropriate evacuation route, participants tended to act with greater confidence and showed less hesitation in their decision-making. In contrast, when all agents chose an inappropriate route, participants' confidence decreased. Although no statistical comparison was conducted, the evacuation success rate also tended to decline as agents increasingly selected inappropriate routes. These findings suggest that participants may rely more on the collective behavior of agents than on their own sense of uncertainty when making decisions during evacuation. This tendency highlights the importance of incorporating psychological factors into disaster-preparedness training.

In Experiment 3, the presence of agents capable of expressing emotion conveyed an impression of fear-driven behavior. This indicates that the emotional expressions implemented in the system were successfully understood by participants. However, the participants' own feelings of fear were not elicited, and no emotional contagion occurred. This is likely attributable to participants becoming habituated to the disaster scenario and the display of fear through repeated trials.

The results of this study suggest that the behaviors of other agents can influence users' decision-making even under crisis conditions.

\subsection{Limitation}
This study has several limitations. First, the VR environment used in the experiments repeatedly employed a single building layout and did not include the spatial complexity or obstacles found in real disaster sites. Therefore, the realism of the disaster experience provided in this study may have been insufficient. In addition, the influence of participants' habituation to the VR-based disaster environment on their fear responses and decision-making was not considered. The attributes of the participants were also limited, and the effects of differences in age or disaster experience on psychological responses during the drill remain unexamined. These factors constrain the generalizability of the findings. Future work should implement disaster environments with higher realism and examine the possibility of misguidance by groups of agents while taking into account factors that may influence users' familiarity with and understanding of the environment.

\subsection{Trust and Blame}
The results of this study suggest that the behaviors of other agents can influence users' decision-making even under crisis conditions. In VR-based disaster evacuation training, if users make incorrect decisions by relying on agent behaviors, they may attribute the failure to the agents, which could hinder the development of autonomous learning. Furthermore, excessive fear expressions or strongly directive behaviors by multiple agents may increase users' anxiety and psychological burden, posing a risk of the training becoming traumatic. To avoid these issues, it is important to design an environment in which users can remain emotionally regulated and objectively reflect on their evacuation behaviors, including their interactions with agents. Therefore, VR disaster training systems and the multi-agent behaviors employed within them should clearly communicate the training objectives beforehand and appropriately disclose the agents' behavioral principles afterward, while providing feedback mechanisms that facilitate calm and constructive reflection.

\section{Conclusion}
This study proposed a disaster preparedness training system that utilizes VR to simulate disaster scenarios and incorporates multiple virtual agents to account for group psychological phenomena, such as conformity bias, with the aim of raising disaster awareness. An investigation into the impact of the number and behavior of virtual agents revealed two key findings: evacuation drills conducted without the presence of other agents induce a greater sense of fear, and the behavior of agents influences decisions regarding evacuation routes. Future work will involve simulating group psychology to examine whether users are emotionally influenced by the propagation of panic or distress among others. Additionally, we plan to link these dynamics to the disaster scenario so that they respond to events such as partial structural collapses or the spread of fire.

\begin{acks}
This research was supported in part by JSPS KAKENHI Grant Numbers 23K11278, 26K14911, 23K11202, and 24K02977.
\end{acks}


\bibliographystyle{ACM-Reference-Format}
\bibliography{references}










\end{document}